\documentclass[12pt]{article}
\usepackage{amsmath}
\usepackage[cp1251]{inputenc}
\usepackage{bm}
\usepackage{color}
\usepackage{caption}
\usepackage{subcaption}
\usepackage{graphicx}
\usepackage{braket}
\begin{document}
\title{Geometric properties of evolutionary graph states and their detection on a quantum computer}

\author{
Kh. P. Gnatenko$^1$\footnote{khrystyna.gnatenko@gmail.com}, H. P. Laba$^2$\footnote{hanna.laba@polynet.lviv.ua},
V. M. Tkachuk$^1$\footnote{voltkachuk@gmail.com}\\
$^1$Professor Ivan Vakarchuk Department for Theoretical Physics,\\
Ivan Franko National University of Lviv,\\
12, Drahomanov St., Lviv, 79005, Ukraine.\\
$^2$Department of Applied Physics and Nanomaterials Science, \\
Lviv Polytechnic National University,\\
5 Ustiyanovych St., 79013 Lviv, Ukraine.}

\maketitle

\begin{abstract}
Geometric properties of evolutionary graph states of spin systems generated by the operator of evolution with Ising Hamiltonian are examined, using their relationship with fluctuations of energy. We find that the geometric characteristics of the graph states depend on properties of the corresponding graphs. Namely, it is obtained that the fluctuations of energy in graph states and therefore the velocity of quantum evolution, the curvature and the torsion of the states are related with the total number of  edges, triangles and squares in the corresponding graphs.  The obtained results give a possibility to quantify the number of  edges, triangles and squares in a graph on a quantum devise and achieve quantum supremacy in solving this problem with the development of a multi-qubit quantum computer.   Geometric characteristics of graph states corresponding to a chain, a triangle, and a square are detected on the basis of calculations on IBM's quantum computer  $\textrm{ibmq\_manila}$.

\end{abstract}

\vspace{0.5cm}

\section{Introduction}

Without any doubt
geometric ideas are important in studies of  problems of quantum information, among them examining of entanglement of quantum states
\cite{Brody01,Brody07,Fry17}, studies of quantum evolution
\cite{Anandan90,Abe93,Grigorenko91,Kuzmak16}, solving quantum brachistochrone problem \cite{Mosta07,Bender08,Fryd08,Kuzmak15}. Distance between quantum states can be used for measure of entanglement. Namely, Abner Shimony
in his paper \cite{Shim95} proposed
the geometric measure of entanglement which is defined as minimal squared Fubiny-Study distance between an entangled state and a set of separable pure states.
The authors of  recent paper \cite{Gir21}
 introduced the weighted distances, namely a
new class of information-theoretic measures that quantify how hard it is
to discriminate between two quantum states of many particles.
It is worth also noting  paper \cite{Xie20} where the first experiment on measuring the geometry
of quantum states in a three-level system was reported.

In the classical motion the curvature and
torsion are important geometric characteristics of the trajectory.
Expression for the curvature of quantum evolution was derived
in \cite{Brody96}. In \cite{Lab17} we found expression for the curvature and torsion for evolution
of quantum system.

In the present paper we study geometric properties of graph states.
It is worth mentioning that  graph states  have been widely studied  (see, for instance, \cite{Wang,Mooney,Hein,Hein1,Guhne,Markham,Cabello,Alba,Mezher,Gna21,Susulovska} and references therein) because of their importance, for instance, in
quantum cryptography \cite{Markham,Qian}, quantum error correction \cite{Schlingemann,Bell,Mazurek}. Recent studies have been devoted to examining entanglement of the graph states on quantum computers \cite{Wang,Mooney,Gna21,Susulovska}.

We consider graph states of spin systems generated by operator of evolution with Ising Hamiltonian. Expressions for the velocity of evolution, the curvature and the torsion are obtained. We show that the velocity of quantum evolution is related with the total number of edges in the graph,
the curvature is related with total number of edges and squares in the graph and the torsion in addition depends on the total number of triangles in the graph. For particular cases of graph states (graph states corresponding to a chain, a triangle, a square) we detect geometric properties of the states in evolution on the basis of calculations on IBM's quantum computer.

The paper is organized as follows. In Section 2 we present relation of geometric characteristics of quantum states with the fluctuations of energy.  Section 3 is devoted to studies of the geometric characteristics of quantum graph states  generated by operator of evolution with Ising Hamiltonian and their relation with the graph properties. In Section 4 we detect geometric characteristic of graph states corresponding to a chain, a triangle, a square on IBM's quantum computer. Conclusions are presented in Section 5.

\section{Velocity, curvature and torsion of quantum states in evolution}

Let us consider the geometric properties of quantum states in evolution.
For a system described by Hamiltonian  $H$  the velocity of quantum
evolution is defined as (see \cite{Anandan90})
\begin{eqnarray} \label{velocity}
v={ds\over dt}=\frac{\gamma\sqrt{\langle(\Delta H)^2\rangle}}{ \hbar},
\end{eqnarray}
 here  $\Delta H=H-\langle H\rangle$ and
\begin{eqnarray}
\langle(\Delta H)^2\rangle=\langle\psi(t)|(\Delta H)^2|\psi(t)\rangle,\\
|\psi(t)\rangle=e^{-iHt/\hbar}|\psi(0)\rangle.
\end{eqnarray}
Note that in the case when a Hamiltonian does not depend explicitly on time the velocity of quantum
evolution is a constant.

The  geodesic line is defined as one-parametric set of the quantum state vectors that connects two state vectors $|\psi_0\rangle$, $|\psi_1\rangle$ with linear combination
\begin{eqnarray}\label{geodesic}
|\psi(\xi)\rangle={1\over\sqrt{1-2\xi(1-\xi)(1-|\langle\psi_1|\psi_0\rangle|)}}\times\nonumber\\ \times\left[(1-\xi)|\psi_0\rangle+\xi|\psi_1\rangle
{\langle\psi_1|\psi_0\rangle\over
|\langle\psi_1|\psi_0\rangle|}\right],
\end{eqnarray}
where   $\xi$ is a real parameter changing from $0$ to $1$ (for the details see \cite{Lab17}).
The length of the geodesic line is equal to the
Wootters distance between the corresponding state vectors
\begin{eqnarray}
s=\int ds=\gamma\arccos|\langle\psi_1|\psi_0\rangle|.
\end{eqnarray}

The curvature of quantum evolution is defined as deviation of evolution state vector $|\psi(t)\rangle$
from the geodesic connecting the two evolutionary states.
For small times one can treat the classical motion along a given curve as a circular motion
with radius $R$. Using notation $s$ for  the length of the curve between two neighboring
points, which can be considered as an arc of the circle, and
$d$ for the  distance between the middle point of an arc and the
chord connecting these two points one can write ${1/R}={8d/(s)^2}$.
The radius of curvature can be rewritten also in the following form
${1/ R^2}=24\left(1-{l/ s}\right)/ s^2$, where $l$ is the geodesic distance between two closed  quantum states in evolution,
$s$ is the length of quantum evolution pass.
Similarly to classical definition,  in quantum case the radius of curvature reads
\begin{eqnarray}
{1\over R^2}={\langle(\Delta
H)^4\rangle-\langle(\Delta H)^2\rangle^2\over\gamma^2\langle(\Delta
H)^2\rangle^2}={\bar \kappa\over\gamma^2},\label{curvvv}
\end{eqnarray}
(for the details see \cite{Lab17}). Here  for convenience we introduce constant
$\bar \kappa=\gamma^2/ R^2$.

The torsion can be defined as deviation of evolution state vector from the plane of evolution
at a given time \cite{Lab17}. The plane of evolution is a two-dimensional subspace spanned by two close  evolutionary states.
The coefficient that characterizes such a deviation is called torsion coefficient and is given by
\begin{eqnarray}
\tau=\langle(\Delta H)^4\rangle-\langle(\Delta
H)^2\rangle^2-{\langle(\Delta H)^3\rangle^2\over\langle(\Delta
H)^2\rangle}.
\end{eqnarray}
 Dimensionless torsion coefficient can be introduced as
\begin{eqnarray}\label{bartor}
\bar\tau={\tau\over\langle(\Delta H)^2 \rangle^2} =
{\langle(\Delta H)^4\rangle-\langle(\Delta
H)^2\rangle^2\over\langle(\Delta H)^2\rangle^2}-\nonumber\\-{\langle(\Delta
H)^3\rangle^2\over\langle(\Delta H)^2\rangle^3}=\bar
\kappa-{\langle(\Delta H)^3\rangle^2\over\langle(\Delta
H)^2\rangle^3}.
\end{eqnarray}

In the next sections on the basis of the relations  we study the geometric properties of the graph states of spin systems generated by operator of evolution with Ising Hamiltonian.

\section{Geometric properties of graph states of spin systems with Ising interaction}

Let us consider a spin system described by  Ising Hamiltonian
\begin{eqnarray}
H={1\over 2}\sum_{i,j}J_{ij}\sigma^x_i\sigma^x_j,\label{ising}
\end{eqnarray}
here $\sigma_i^x$ is the Pauli matrix of spin $i$,  $J_{ij}$  is the interaction coupling ($J_{ii}=0$), $i,j=1..N$, $N$ is the number of spins. We consider $J_{ij}=J$, if  the interaction between spin $i$ and spin $j$ exists. Interaction coupling constants $J_{ij}$ can be related with the elements of  adjacency matrix  $A_{ij}$  of an undirected graph as $J_{ij}=A_{ij}J$.
Therefore the evolutionary state
\begin{eqnarray}
\vert\psi\rangle=e^{-\frac{it}{2\hbar}\sum_{i,j}J_{ij}\sigma_i^x\sigma_j^x}\vert\psi_0\rangle,\label{state}\\
\vert\psi_0\rangle=\vert00...0\rangle,\label{zero}
\end{eqnarray}
is a graph state with vertices represented by the spins and edges corresponding to the interactions between them.  Note, that the spin states correspond to qubit states,  $\ket{\uparrow}=\ket{0}$,  and  $\ket{\downarrow}=\ket{1}$.

As was obtained in \cite{Lab17} and presented in the previous section, to find geometric properties of quantum graph states it is necessary to calculate the mean values $\braket{\Delta H^2}$, $\braket{\Delta H^3}$, $\braket{\Delta H^4}$.  Note that Hamiltonian (\ref{ising}) does not depend on time. As a result $\braket{\Delta H^2}$, $\braket{\Delta H^3}$, $\braket{\Delta H^4}$ do not depend on time too.
So, for simplicity we consider $t=0$.

For the mean value of Hamiltonian (\ref{ising}) we have
\begin{eqnarray}
\langle H\rangle=
{1\over 2}\sum_{i,j}J_{ij}\langle 00..0|\sigma^x_i\sigma^x_j|00...0\rangle=0,
\end{eqnarray}
where we take into account that $\sigma^x|0\rangle=|1\rangle$ and $i\neq j$.
Squared fluctuation of energy reads
\begin{eqnarray}
\langle \Delta H^2\rangle=\nonumber\\
={1\over 4}\sum_{i_1,j_1}\sum_{i_2,j_2}J_{i_1j_1}J_{i_2j_2}\langle 00..0|
\sigma^x_{i_1}\sigma^x_{j_1}\sigma^x_{i_2}\sigma^x_{j_2}|00...0\rangle.
\end{eqnarray}
In this sum we obtain nonzero term if $i_1=i_2$ and $j_1=j_2$ or
$i_1=j_2$ and $j_1=i_2$.
Then we can write
\begin{eqnarray}
\langle \Delta H^2\rangle=
{1\over 2}\sum_{i,j}J^2_{ij}=k_2J^2,\label{dh2}
\end{eqnarray}
where $k_2$ is the total number of edges in the graph.

Let us  also calculate $\langle  \Delta H^3\rangle$. We have
\begin{eqnarray}
\langle \Delta H^3\rangle=\langle H^3\rangle=\sum_{i_1>j_1}\sum_{i_2>j_2}\sum_{i_3>j_3}
J_{i_1j_1}J_{i_2j_2}J_{i_3j_3}\times\nonumber\\ \times \langle 00..0|
\sigma^x_{i_1}\sigma^x_{j_1}\sigma^x_{i_2}\sigma^x_{j_2}\sigma^x_{i_3}\sigma^x_{j_3}|00...0\rangle.
\end{eqnarray}
Each therm in this sum gives nonzero contribution only in the case when
three edges create a triangle.
We find
\begin{eqnarray}
\langle\Delta H^3\rangle=3!J^3k_3=6J^3k_3,\label{dh3}
\end{eqnarray}
where $k_3$ is the total number of triangles in the graph, the multiplier $3!$ is the number of combinations of three edges.
For $\langle\Delta H^4\rangle$ we obtain
\begin{eqnarray}
\langle\Delta H^4\rangle=\langle H^4\rangle=
J^4\left(k_2+3k_2(k_2-1)+4!k_4\right),\label{dh4}
\end{eqnarray}
here
$k_4$ is the total number of squares in graph, multiplier $4!$ is the number of combinations of four edges.

Taking into account (\ref{velocity}) and using (\ref{dh2}), we obtain the velocity of evolution of the graph state as
\begin{eqnarray}
v={\gamma J\over \hbar}\sqrt{k_2}.\label{rv}
\end{eqnarray}
Substituting obtained results (\ref{dh2}), (\ref{dh3}), (\ref{dh4}) into expressions for curvature (\ref{curvvv}) and torsion (\ref{bartor})   we have
\begin{eqnarray}
{\gamma^2\over R^2}={\bar\kappa}=\frac{1}{k_2^2}\left(k_2+3k_2(k_2-1)+4!k_4\right)^2-1,\label{rc}\\
\bar\tau={\bar\kappa}-{6^2k_3^2\over k_2^3}.\label{rt}
\end{eqnarray}

It is important to note that we obtain that  velocity of quantum evolution  (\ref{rv}) is related with the total number of links in a graph, curvature (\ref{rc}) is related with the total number of links and squares in the graph and torsion (\ref{rt})  in addition depends on the total number of triangles in the graph. So, there is relation of the geometric properties of evolutionary graph states with the graph properties.

In the next section we present quantum protocols for detection of the geometric properties of evolutionary graph states  on a quantum device and results of realization of the protocols on IBM's quantum computer  $\textrm{ibmq\_manila}$.

\section{Detecting geometric properties of graph states on IBM's quantum computer}

Let us calculate geometric properties of quantum graph states on a quantum computer. For this purpose we use relations of the properties with the fluctuations of energy obtained in \cite{Lab17}. As examples we study quantum graph states corresponding to a chain, a triangle and a square on  $\textrm{ibmq\_manila}$ \cite{kk}.

\subsection{Graph state corresponding to a chain}
Let us consider a chain of three spins with  Ising interaction, described by the following Hamiltonian
\begin{eqnarray}
H=J\sigma^x_0\sigma^x_1+J\sigma^x_1\sigma^x_2,\label{shams2}
\end{eqnarray}
where $J$ is the interaction coupling constant. In this case  quantum graph state reads
\begin{eqnarray}
\ket{\psi}=e^{iJ(\sigma^x_0\sigma^x_1+\sigma^x_1\sigma^x_2)/\hbar}\ket{000}.\label{grchain}
\end{eqnarray}

To calculate squared fluctuation of energy on  a quantum device  we study  the mean value of the evolution operator. For small times it can be written as
\begin{eqnarray}
U=\bra{0..0}e^{iHt/\hbar}\ket{0...0}=1-\frac{\braket{\Delta H^2}}{2\hbar^2}t^2.
\end{eqnarray}
Then for $|U|^2$ we obtain
\begin{eqnarray}\label{rel}
|U|^2=1-\frac{\braket{\Delta H^2}}{\hbar^2}t^2.
\end{eqnarray}
The value $|U|^2$ can be detected on a quantum computer as a function of time. Then on the basis of this result  and relation (\ref{rel}) one can find the squared fluctuations of energy.

\begin{figure}[!!h]
\begin{center}
{\includegraphics[scale=0.6, angle=0.0, clip]{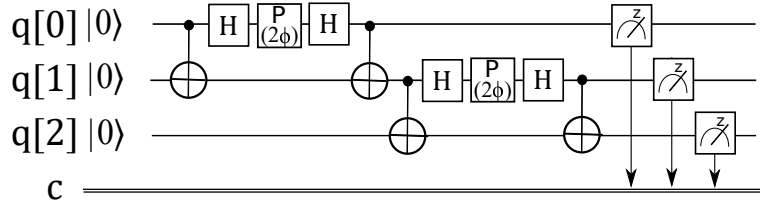}}
\end{center}
\caption{Quantum protocol for detecting $|U|^2=|\bra{000}e^{iHt/\hbar}\ket{000}|^2$ in the case of spin chain (\ref{shams2}), $\phi=Jt/\hbar$.}
		\label{sfig:1}
\end{figure}

Quantum protocol for studies of the mean value of operator of evolution in the case of spin chain (\ref{shams2})  is presented in Fig. \ref{sfig:1}.

\begin{figure}[!!h]
\begin{center}
{\includegraphics[scale=0.35, angle=0.0, clip]{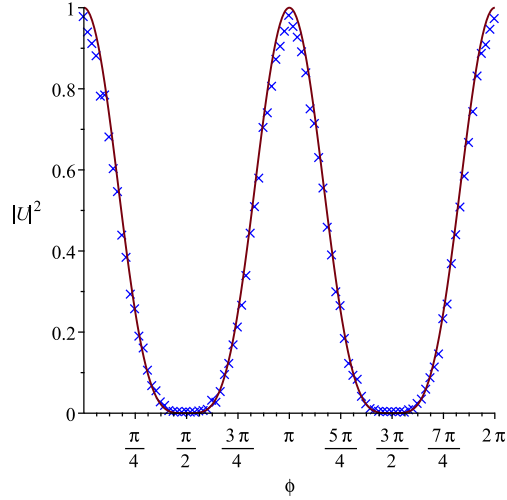}}
\end{center}
\caption{Results for $|U|^2=|\bra{000}e^{iHt/\hbar}\ket{000}|^2$ for different values of $\phi=Jt/\hbar$ in the case of  spin chain (\ref{shams2}) obtained  on $\textrm{ibmq\_manila}$ (crosses) and analytical ones (line). }
		\label{sfig:2}
\end{figure}

\begin{figure}[!!h]
\begin{center}
{\includegraphics[scale=0.35, angle=0.0, clip]{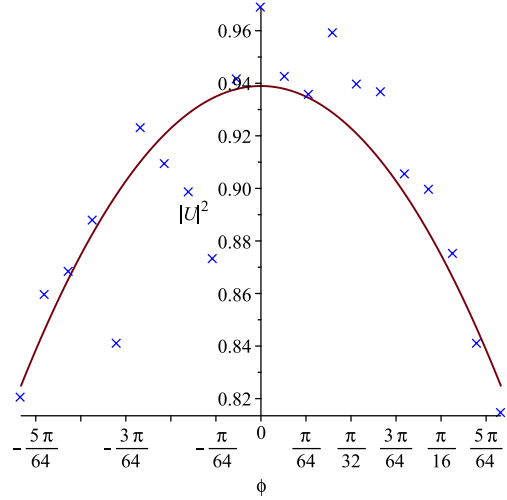}}
\end{center}
\caption{Results of detecting $|U|^2=|\bra{000}e^{iHt/\hbar}\ket{000}|^2$ on $\textrm{ibmq\_manila}$  for different values of $\phi=Jt/\hbar$ close to zero in the case of the spin chain (\ref{shams2}) (marked by crosses) and fitting curve $-1.67\phi^2+0.94$ (line).}
		\label{sfig:3}
\end{figure}

As a result of action of gates $CX_{01}H_0P_0(2\phi)\times$ $\times H_0CX_{01}CX_{12}H_1P_1(2\phi)H_1CX_{12}$ on $\ket{000}$ graph state (\ref{grchain}) is prepared, here $\phi=Jt/\hbar$.
Here $CX_{ij}$ is the controlled-NOT gate acting on qubit $q[i]$
as control and on $q[j]$ as target, $H_i$ is the Hadamard gate.
 On the basis of the results of measurements in the standard basis we obtain
\begin{eqnarray}
|U|^2=|\langle{000}|{\psi}\rangle|^2.
\end{eqnarray}
Quantum protocol Fig. \ref{sfig:1} was realized on  $\textrm{ibmq\_manila}$ for different moments of time. Namely changing $\phi=Jt/\hbar$  from $0$ to $2\pi$ with  step $\pi/48$ we detect  dependence of  $|U|^2$ on time. The results of quantum calculations are presented in  Fig. \ref{sfig:2}.
In order to detect $\braket{\Delta H^2}$ we studied  $|U|^2$ close to $t=0$. Namely, quantum protocol Fig. \ref{sfig:1} was realized on  $\textrm{ibmq\_manila}$ for $\varphi=Jt/\hbar$  in range from $-\pi/12$ to $\pi/12$ changing with step $\pi/120$. In this case, taking into account (\ref{rel}), one can fit the the obtained result by $-a\phi^2+b$ ($a$, $b$ are  constants) and find  $\braket{\Delta H^2}=aJ^2$.
   Note that  $\braket{\Delta H}=0$, so $\braket{\Delta H^2}=\braket{H^2}=aJ^2$.
   The results of calculations on the quantum device and results of fitting by the least squares  are presented in Fig. \ref{sfig:3}.
   We find $\braket{H^2}=1.67J^2$. Note that the obtained result is close to the analytical one, which is $\braket{H^2}=2J^2$.

\begin{figure}[!!h]
\begin{center}
{\includegraphics[scale=0.65, angle=0.0, clip]{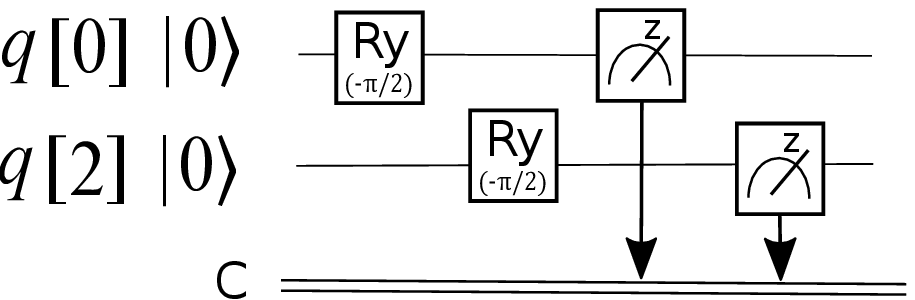}}
\end{center}
\caption{Quantum protocol for detecting $\braket{\sigma^x_0\sigma^x_2}$.}
		\label{sfig:4}
\end{figure}

It is worth also mentioning that
\begin{eqnarray}
H^2=(J\sigma^x_0\sigma^x_1+J\sigma^x_1\sigma^x_2)^2=2J^2(1+\sigma^x_0\sigma^x_2).\label{sihh}
\end{eqnarray}
So, the value of $\braket{H^2}$ can be found detecting $\braket{\sigma^x_0\sigma^x_2}=\bra{000}\sigma^x_0\sigma^x_2\ket{000}$ on the quantum devise. The quantum protocol for such studies is presented in Fig. \ref{sfig:4}.
In the protocol  we take into account that operator $\sigma^x$ can be represented as $\sigma^x=\exp(-i\pi\sigma^y/4)\sigma^z\exp(i\pi\sigma^y/4)$.  So, we can write
\begin{eqnarray}
 \bra{000}\sigma^x_0\sigma^x_2\ket{000}=\nonumber\\=\bra{00}e^{-i\pi\sigma_0^y/4}\sigma^z_0e^{i\pi\sigma_0^y/4}e^{-i\pi\sigma_2^y/4}\sigma^z_2e^{i\pi\sigma_2^y/4}\ket{00}=\nonumber\\
 =\bra{\tilde{\psi}}\sigma^z_0\sigma^z_2\ket{\tilde{\psi}}=|\langle\tilde{\psi}|{00}\rangle|^2-|\langle\tilde{\psi}|{10}\rangle|^2-\nonumber\\-|\langle\tilde{\psi}|{01}\rangle|^2+|\langle\tilde{\psi}|{11}\rangle|^2,
 \end{eqnarray}
where $\ket{00}=\ket{0}_0\ket{0}_2$ and
\begin{eqnarray}
\ket{\tilde{\psi}}=e^{i\pi\sigma_0^y/4}e^{-i\pi\sigma_2^y/4}\ket{00}.
 \end{eqnarray}
 The value $\langle\sigma^x_0\sigma^x_2\rangle$ can be calculated using  results of measurements of states of qubits $q[0]$, $q[2]$ after their rotation by $\pi/2$ around the $y$ axis,  see Fig. \ref{sfig:4}.
On the basis of the results of measurements we find $\braket{\sigma^x_0\sigma^x_2}=-0.0117$. So, taking into account (\ref{sihh}), we obtain
\begin{eqnarray}
\braket{H^2}=2J^2(1+\braket{\sigma^x_0\sigma^x_2})=1.98J^2.\label{hh}
\end{eqnarray}
The result is close to the analytical one

Similarly, on the quantum computer we calculate $\braket{\sigma^x_0\sigma^x_1}=0.014$, $\braket{\sigma^x_1\sigma^x_2}=-0.027$ and obtain
\begin{eqnarray}
\braket{H^3}=J^3(\braket{\sigma^x_0\sigma^x_1}+\braket{\sigma^x_0\sigma^x_1})=-0.013J^3,\label{qhh}\\
\braket{H^4}=7.92J^4.\label{qqhh}
\end{eqnarray}
The results of quantum calculations  are close to that obtained analytically $\braket{H^3}=J^2\braket{H}=0$,  $\braket{H^4}=4J^2\braket{H^2}=8J^4$.

Finding fluctuations of energy, one can also detect curvature and torsion. We have
\begin{eqnarray}
\frac{\gamma^2}{R^2}=\bar{\tau}=1.02.\label{rrrrrr}
\end{eqnarray}
Theoretical result for these values is ${\gamma^2}/{R^2}=\bar{\tau}=1$.

On the basis of outcomes of quantum calculations (\ref{hh})-(\ref{qqhh}) and relations (\ref{dh2}), (\ref{dh3}), (\ref{dh4}), rounding to the nearest integer number,  we find  $k_2=2$, $k_3=k_4=0$ that correspond to the number of edges, triangles and squares in the chain graph with three nodes.

\subsection{Graph state corresponding to a triangle}
Let us consider a spin system with the following Hamiltonian
\begin{eqnarray}
H=J\sigma^x_0\sigma^x_1+J\sigma^x_1\sigma^x_2+J\sigma^x_0\sigma^x_2.\label{shams3}
\end{eqnarray}
Starting from $\ket{000}$ as a result of evolution one obtains the state
\begin{eqnarray}
\ket{\psi}=e^{iJ(\sigma^x_0\sigma^x_1+\sigma^x_1\sigma^x_2+\sigma^x_2\sigma^x_0)/\hbar}\ket{000}.\label{grtr}
\end{eqnarray}

\begin{figure}[!!h]
\begin{center}
{\includegraphics[scale=0.55, angle=0.0, clip]{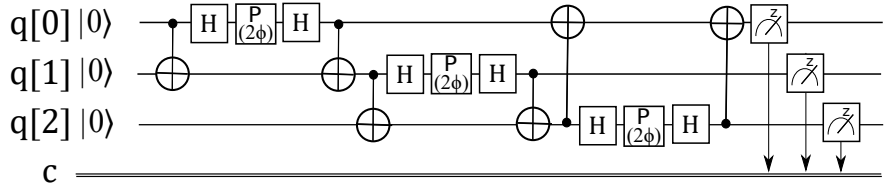}}
\end{center}
\caption{Quantum protocol for detecting $|U|^2=|\bra{000}e^{iHt/\hbar}\ket{000}|^2$ in the case of  spin triangle (\ref{shams3}), $\phi=Jt/\hbar$.}
		\label{sfig:5}
\end{figure}

Similarly as in the previous example we study the mean value of operator of evolution  on the quantum computer.
Quantum protocol for such studies is presented in Fig. \ref{sfig:5}.

\begin{figure}[!!h]
\begin{center}
\subcaptionbox{\label{ff1}}{\includegraphics[scale=0.3, angle=0.0, clip]{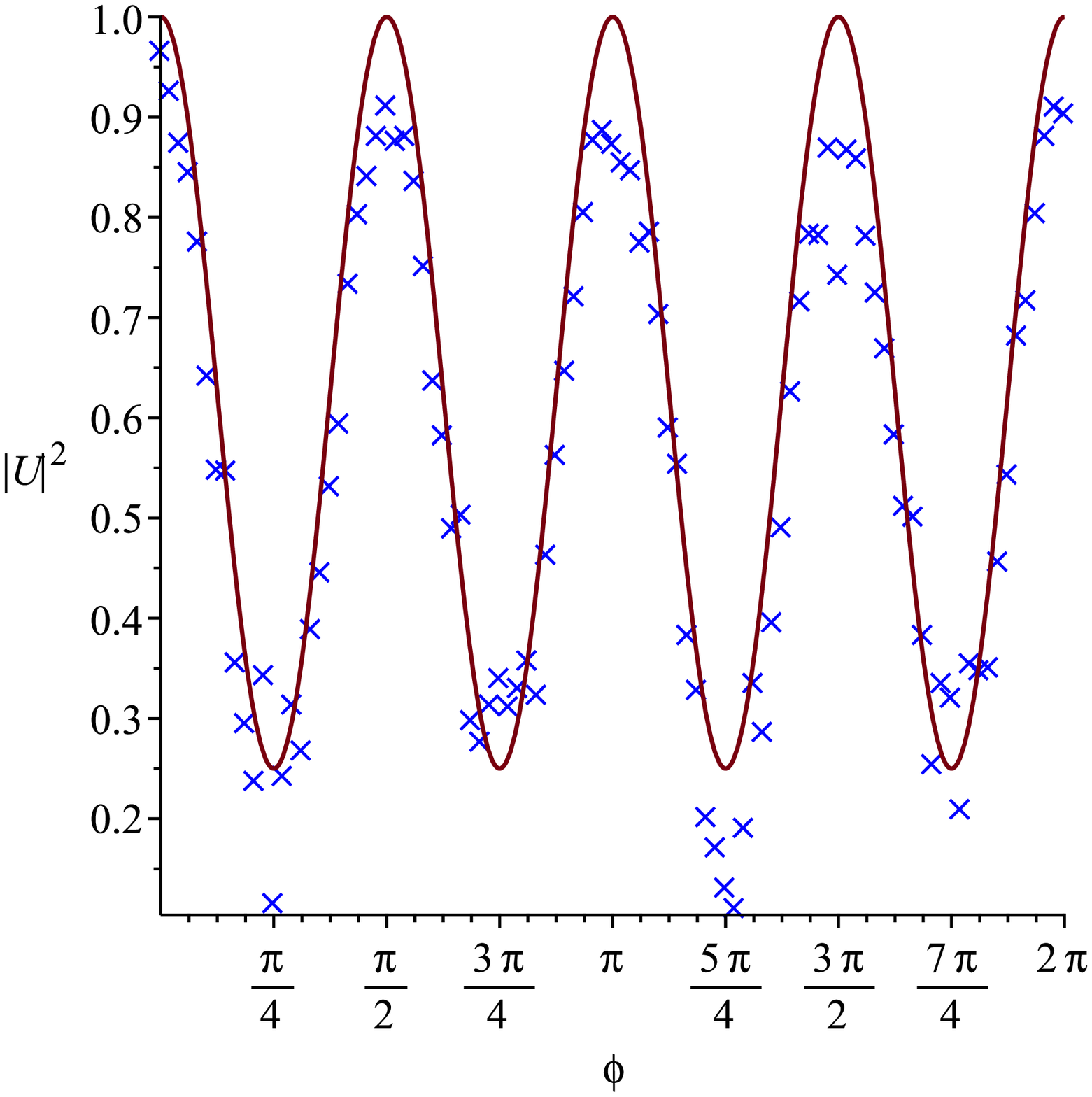}}
\hspace{1cm}
\subcaptionbox{\label{ff3}}{\includegraphics[scale=0.3, angle=0.0, clip]{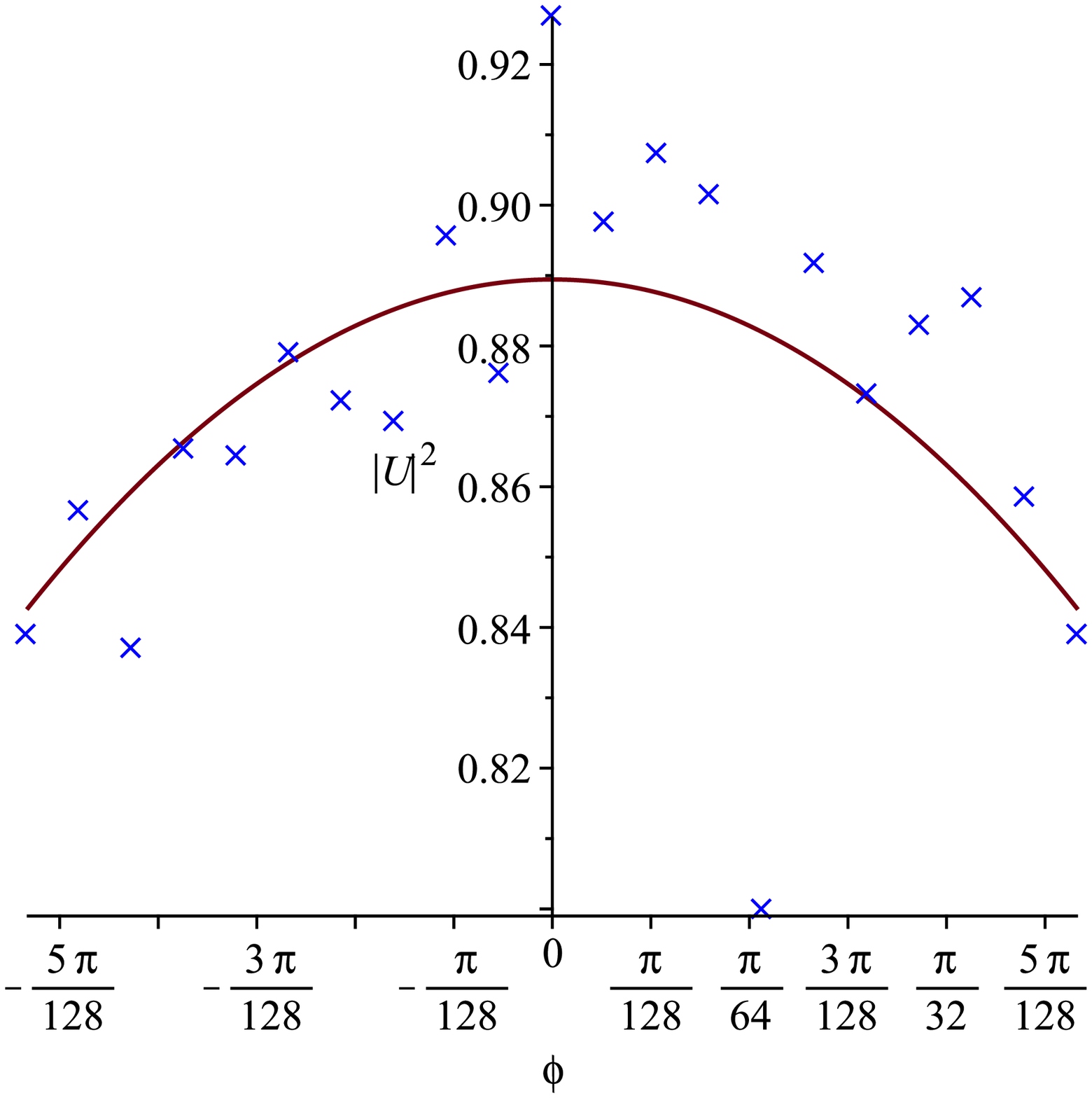}}
\end{center}
\caption{Results for $|U|^2=|\bra{000}e^{iHt/\hbar}\ket{000}|^2$ for different values of $\phi=Jt/\hbar$ in the case of spin triangle (\ref{shams3}) obtained on $\textrm{ibmq\_manila}$ (crosses), analytical ones (line (a)) and fitting curve $-2.74 \phi^2+0.89$ (line (b)). }
		\label{sfig:6}
\end{figure}

We realized  quantum protocol Fig. \ref{sfig:5}  on  $\textrm{ibmq\_manila}$.  Parameter $\phi=Jt/\hbar$  was changed from $0$ to $2\pi$ with the step $\pi/48$. The results  are presented in  Fig. \ref{sfig:6} (a). Also,  the value of $|U|^2$ was quantified for $\phi$ close to zero. Namely,  the  parameter $\phi$  was changed from $-\pi/24$ to $\pi/24$ with the step $\pi/240$, see  Fig. \ref{sfig:6} (b).
The obtained results where fitted by $-a\phi^2+b$. We found  $\braket{\Delta H^2}=2.74J^2$ that is close to that obtained on the basis of analytical calculations.
\begin{eqnarray}
\braket{\Delta H^2}=J^2(3+2\braket{\sigma^x_0\sigma^x_1}+2\braket{\sigma^x_1\sigma^x_2}+2\braket{\sigma^x_0\sigma^x_2})=\nonumber\\=3J^2.
\end{eqnarray}

Another way to quantify the fluctuations of the energy in  graph state corresponding to a triangle (\ref{grtr}) is to
 detect the mean values $\braket{\sigma^x_0\sigma^x_1}$, $\braket{\sigma^x_1\sigma^x_2}$, $\braket{\sigma^x_0\sigma^x_2}$.
Similarly as in the previous subsection, we quantify $\braket{\sigma^x_0\sigma^x_1}$, $\braket{\sigma^x_1\sigma^x_2}$, $\braket{\sigma^x_0\sigma^x_2}$ in  state $\ket{000}$  on quantum device  $\textrm{ibmq\_manila}$ and obtain
\begin{eqnarray}
\braket{\Delta  H^2}=2.95J^2.\label{3hh}\\
\braket{\Delta  H^3}=J^3(7\braket{\sigma^x_0\sigma^x_1}+7\braket{\sigma^x_1\sigma^x_2}+\nonumber\\+7\braket{\sigma^x_0\sigma^x_2}+6)=5.83J^3,\label{33hh}\\
\braket{\Delta H^4}=J^4(20\braket{\sigma^x_0\sigma^x_1}+20\braket{\sigma^x_1\sigma^x_2}+\nonumber\\+20\braket{\sigma^x_0\sigma^x_2}+21)=20.5J^4.\label{333hh}
\end{eqnarray}
On the basis of results for  fluctuations of the energy the curvature and the torsion read
\begin{eqnarray}
\frac{\gamma^2}{R^2}=1.33,\ \ \bar{\tau}=0.031.
\end{eqnarray}
Note that the results correspond to the theoretical ones  ${\gamma^2}/{R^2}=1.33$ and $\bar{\tau}=0$.

 Using (\ref{dh2})-(\ref{dh4}), (\ref{3hh})-(\ref{333hh}),  and rounding to the nearest integer number,  we find  the number of edges $k_2=3$, triangles  $k_3=1$ and squares $k_4=0$ in the triangle graph.

\subsection{Graph state corresponding to a square graph}
\begin{figure}[!!h]
\begin{center}
{\includegraphics[scale=0.55, angle=0.0, clip]{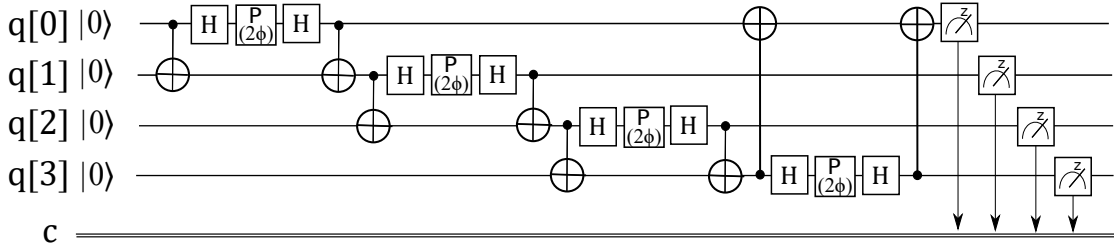}}
\end{center}
\caption{Quantum protocol for detecting $|U|^2=|\bra{0000}e^{iHt/\hbar}\ket{0000}|^2$ in the case of  spin square (\ref{shams5}), $\phi=Jt/\hbar$.}
		\label{sfig:7}
\end{figure}

For a spin system with  Hamiltonian
\begin{eqnarray}
H=J\sigma^x_0\sigma^x_1+J\sigma^x_1\sigma^x_2+J\sigma^x_2\sigma^x_3+J\sigma^x_0\sigma^x_3,\label{shams5}
\end{eqnarray}
 as a result of evolution one obtains the following graph state
\begin{eqnarray}
\ket{\psi}=e^{iJ(\sigma^x_0\sigma^x_1+\sigma^x_1\sigma^x_2+\sigma^x_2\sigma^x_0)/\hbar}\ket{0000},\label{grsq}
\end{eqnarray}
 corresponding to a square graph, where $\ket{0000}$ is the initial state.
Quantum protocol for detecting the mean value of operator of evolution in this case is presented in Fig. \ref{sfig:7}.

\begin{figure}[!!h]
\begin{center}
\subcaptionbox{\label{ff1}}{\includegraphics[scale=0.3, angle=0.0, clip]{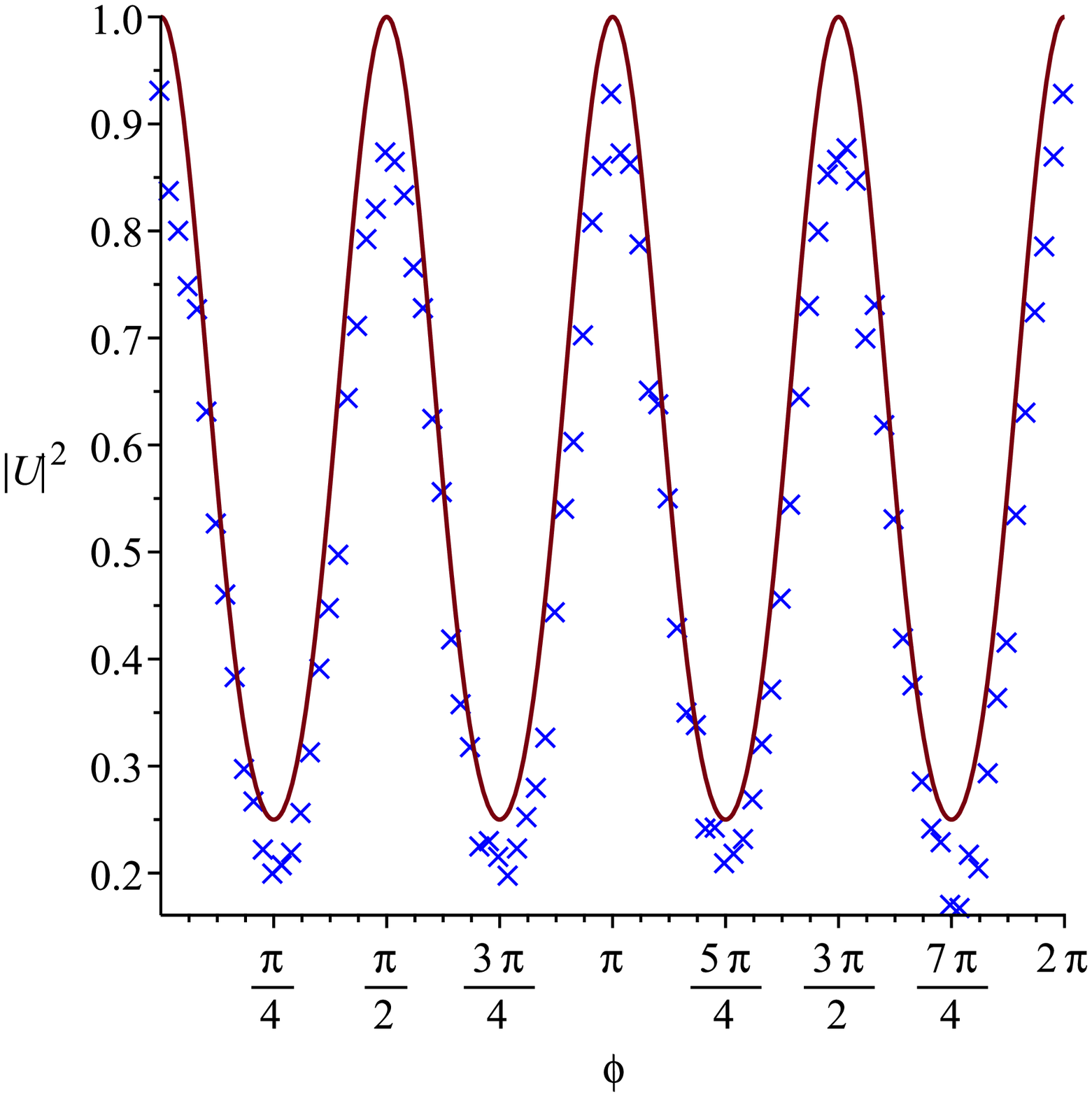}}
\hspace{1cm}
\subcaptionbox{\label{ff3}}{\includegraphics[scale=0.3, angle=0.0, clip]{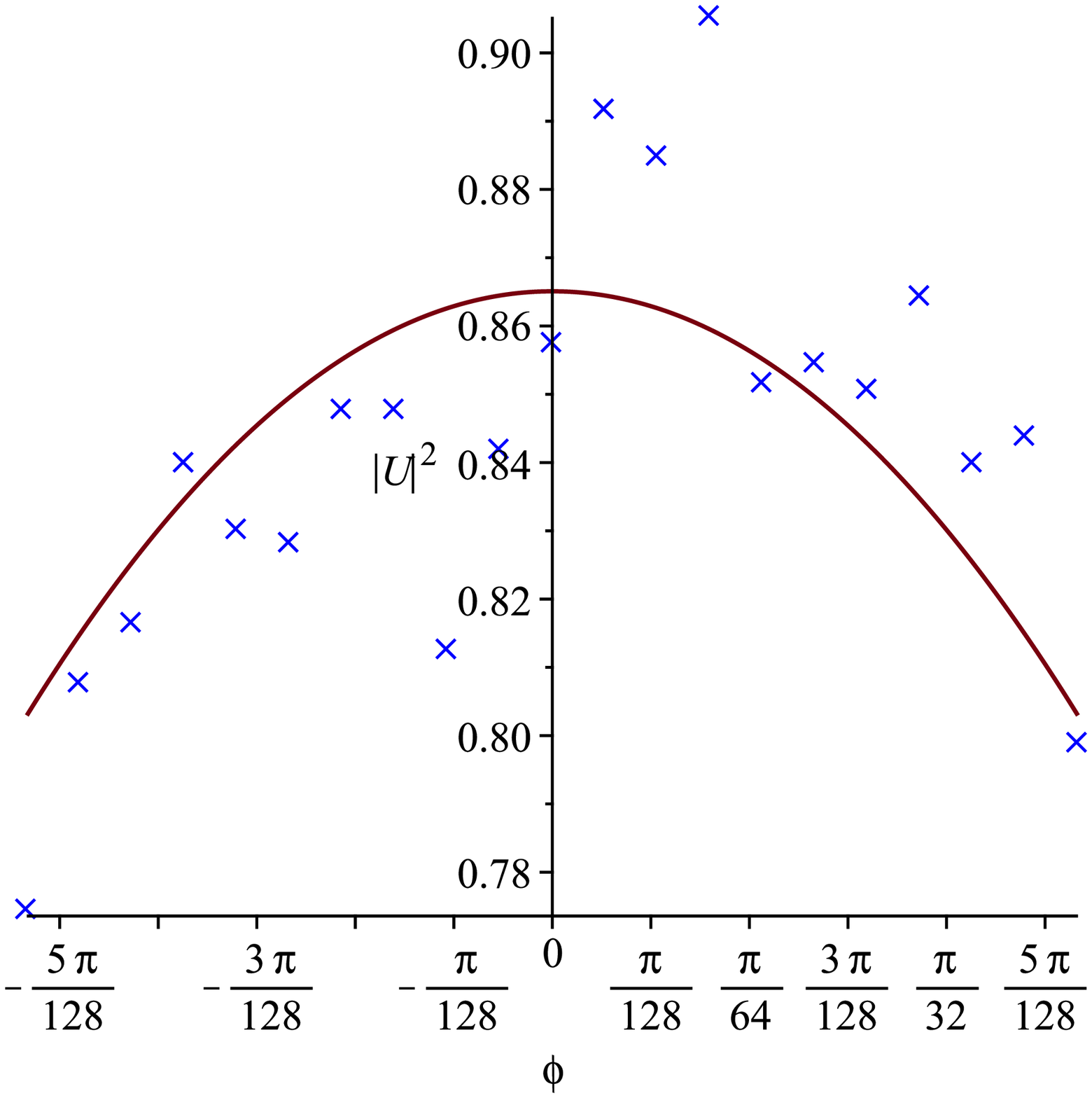}}
\end{center}
\caption{Results for $|U|^2=|\bra{0000}e^{iHt/\hbar}\ket{0000}|^2$  for different values of $\phi=Jt/\hbar$ in the case of the spin square (\ref{shams5}) obtained on $\textrm{ibmq\_manila}$ (crosses), analytical ones (line (a)) and fitting curve $-3.63 \phi^2+0.87$ (line (b)). }
		\label{sfig:8}
\end{figure}

 The  protocol was realized  on  $\textrm{ibmq\_manila}$ for $\phi=Jt/\hbar$ changing from $0$ to $2\pi$ with the step $\pi/48$ and also for $\phi=Jt/\hbar$  changing with the step $\pi/240$ in range from $-\pi/24$ to $\pi/24$ see Fig. \ref{sfig:8} (a), (b).   The results for $|U|^2$  were fitted by $-a\phi^2+b$ (see line in Fig. \ref{sfig:8} (b)). We found  $\braket{\Delta H^2}=3.63J^2$.

Calculating mean values $\braket{\sigma^x_i\sigma^x_j}$ ($i,j=(0,1,2,3)$, $i\neq j$) $\braket{\sigma^x_0\sigma^x_1\sigma^x_2\sigma^x_3}$ in  state $\ket{0000}$ on  $\textrm{ibmq\_manila}$,  we find
\begin{eqnarray}
\braket{\Delta H^2}=J^2(4+4\braket{\sigma^x_0\sigma^x_2}+4\braket{\sigma^x_1\sigma^x_3}+\nonumber\\+4\braket{\sigma^x_0\sigma^x_1\sigma^x_2\sigma^x_3})=3.77J^2,\label{sshh11}\\
\braket{\Delta H^3}=16J^3(\braket{\sigma^x_0\sigma^x_1}+\braket{\sigma^x_1\sigma^x_2}+\braket{\sigma^x_2\sigma^x_3}+\nonumber\\+\braket{\sigma^x_3\sigma^x_0})=1.39J^3,\label{sshh12}\\
\braket{\Delta  H^4}=16\braket{H^2}=60.32J^4.\label{sshh13}
\end{eqnarray}
Therefore, the curvature and the torsion of evolutionary graph state corresponding to a square graph read
\begin{eqnarray}
\frac{\gamma^2}{R^2}=3.24,\ \ \bar{\tau}=3.21.
\end{eqnarray}
The results are in agreement with  theoretical one ${\gamma^2}/{R^2}=\bar{\tau}=3$.

 On the basis of results of quantum calculations (\ref{sshh11})-(\ref{sshh13}) and analytical ones (\ref{dh2})-(\ref{dh4}), rounding to the nearest integer number,   we have the number of edges $k_2=4$, triangles $k_3=0$ and squares $k_4=1$  in the square graph.

\section{Conclusion}

Geometric properties of evolutionary graph states of spin systems with Ising interaction have been studied. Expressions for the velocity, the curvature and the torsion have been obtained (\ref{rv}), (\ref{rc}), (\ref{rt}).
We have found that the  fluctuations of energy in the graph states and the geometric properties of the  states are related with the number of
edges, triangles and squares in the corresponding graphs.
Namely, $\braket{\Delta H^2}$ is related with the total number of edges  (\ref{dh2}),  $\braket{\Delta H^3}$ is related with the total number of triangles (\ref{dh3}), $\braket{\Delta H^4}$ depends on the  total number of edges and squares (\ref{dh4}).
As a result the velocity of quantum evolution of the graph states is related  with the total number of edges  in the graph (\ref{rv}). The curvature of the evolutionary graph states  depends on the total number of edges and the number of squares in the corresponding  graph (\ref{rc}).  The  torsion is related with the number of edges,  squares and triangles (\ref{rt}).
The obtained results give a possibility to detect the number of triangles, number of squares in graphs on a quantum computer. They also opens  a possibility to achieve a quantum supremacy in studies of the properties of large graphs with development of multi-qubit quantum computer.

  Particular cases of the graph states  corresponding to a chain, a triangle and a square have been considered, We have examined the geometric properties of the states, detecting mean values  $\braket{\Delta H^2}$, $\braket{\Delta H^3}$, $\braket{\Delta H^4}$ on the basis of quantum calculations on  IBM's quantum computer $\textrm{ibmq\_manila}$.   The results of calculations on the quantum device are in agreement with the theoretical ones.

\section*{Acknowledgment}
This work was supported by Project 2020.02/0196  (No. 0120U104801) from National Research Foundation of Ukraine.

\end{document}